%% file: paper.tex
\documentclass[sigplan,screen]{acmart}

\newcommand{\asplossubmissionnumber}{114}

\usepackage[normalem]{ulem}
\usepackage{mathtools}
\usepackage{algorithmic}
\usepackage{algorithm2e}
\usepackage{graphicx}
\usepackage{textcomp}
\usepackage{xcolor}
\usepackage{cleveref}

\usepackage{caption}
\usepackage{subfig}
\usepackage{tablefootnote}
\usepackage{multirow}
\usepackage{xspace}
\usepackage{soul}

\AtBeginDocument{%
  \providecommand\BibTeX{{%
    \normalfont B\kern-0.5em{\scshape i\kern-0.25em b}\kern-0.8em\TeX}}}

\copyrightyear{2024}
\acmYear{2024}
\setcopyright{rightsretained}
\acmConference[ASPLOS '24]{28th ACM International Conference on
Architectural Support for Programming Languages and Operating Systems,
Volume 1}{April 27-May 1, 2024}{La Jolla, CA, USA}
\acmBooktitle{28th ACM International Conference on Architectural Support for
Programming Languages and Operating Systems, Volume 1 (ASPLOS '24), April
27-May 1, 2024, La Jolla, CA, USA}
\acmDOI{10.1145/3617232.3624860}
\acmISBN{979-8-4007-0372-0/24/04}

\begin{document}


\title[Exploiting Human Color Discrimination for Memory- and Energy-Efficient Image Encoding in VR]{Exploiting Human Color Discrimination for Memory- and Energy-Efficient Image Encoding in Virtual Reality}

\author{Nisarg Ujjainkar}
\email{nujjaink@ur.rochester.edu}
\affiliation{%
  \institution{University of Rochester}
  \city{Rochester}
  \state{NY}
  \country{USA}
}

\author{Ethan Shahan}
\email{eshahan@u.rochester.edu}
\affiliation{%
  \institution{University of Rochester}
  \city{Rochester}
  \state{NY}
  \country{USA}
}

\author{Kenneth Chen}
\email{kennychen@nyu.edu}
\affiliation{%
  \institution{New York University}
  \city{New York}
  \state{NY}
  \country{USA}
}

\author{Budmonde Duinkharjav}
\email{budmonde@nyu.edu}
\affiliation{%
  \institution{New York University}
  \city{New York}
  \state{NY}
  \country{USA}
}

\author{Qi Sun}
\email{qisun@nyu.edu}
\affiliation{%
  \institution{New York University}
  \city{New York}
  \state{NY}
  \country{USA}
}

\author{Yuhao Zhu}
\email{yzhu@rochester.edu}
\affiliation{%
  \institution{University of Rochester}
  \city{Rochester}
  \state{NY}
  \country{USA}
}

\date{}

\input{abstract}
\maketitle

\input{macros}
\graphicspath{{figs/}}

\thispagestyle{empty}

\keywords{Virtual Reality, Human Color Perception, Frame Buffer Compression, Accelerator}

\input{sec_intro}

\input{sec_brw}

\input{sec_algo}

\input{sec_hw}

\input{sec_es}

\input{sec_eval}

\input{sec_related}

\input{sec_conc}

\bibliographystyle{plain}
\bibliography{refs}

\end{document}

%% file: abstract.tex
\begin{abstract}

Virtual Reality (VR) has the potential of becoming the next ubiquitous computing platform. Continued progress in the burgeoning field of VR depends critically on an efficient computing substrate.
In particular, DRAM access energy is known to contribute to a significant portion of system energy.
Today's framebuffer compression system alleviates the DRAM traffic by using a numerically lossless compression algorithm.
Being numerically lossless, however, is unnecessary to preserve perceptual quality for humans.
This paper proposes a perceptually lossless, but numerically lossy, system to compress DRAM traffic.
Our idea builds on top of long-established psychophysical studies that show that humans cannot discriminate colors that are close to each other. The discrimination ability becomes even weaker (i.e., more colors are perceptually indistinguishable) in our peripheral vision.
Leveraging the color discrimination (in)ability, we propose an algorithm that adjusts pixel colors to minimize the bit encoding cost without introducing visible artifacts.
The algorithm is coupled with lightweight architectural support that, in real-time, reduces the DRAM traffic by 66.9\%  and outperforms existing framebuffer compression mechanisms by up to 20.4\%.
Psychophysical studies on human participants show that our system introduce little to no perceptual fidelity degradation.

\end{abstract}

%% file: macros.tex

\newcommand{\website}[1]{{\tt #1}}
\newcommand{\program}[1]{{\tt #1}}
\newcommand{\benchmark}[1]{{\it #1}}
\newcommand{\fixme}[1]{{\textcolor{red}{\textit{#1}}}}
\newcommand{\fixed}[1]{{\textcolor{orange}{\textit{#1}}}}

\newcommand*\circled[2]{\tikz[baseline=(char.base)]{
            \node[shape=circle,fill=black,inner sep=1pt] (char) {\textcolor{#1}{{\footnotesize #2}}};}}

\ifx\figurename\undefined \def\figurename{Figure}\fi
\renewcommand{\figurename}{Fig.}
\renewcommand{\paragraph}[1]{\textbf{#1} }
\newcommand{\figline}{{\vspace*{.05in}\hline}}

\newcommand{\Sect}[1]{Sec.~\ref{#1}}
\newcommand{\Fig}[1]{Fig.~\ref{#1}}
\newcommand{\Tbl}[1]{Tbl.~\ref{#1}}
\newcommand{\Equ}[1]{Equ.~\ref{#1}}
\newcommand{\Apx}[1]{Apdx.~\ref{#1}}
\newcommand{\Alg}[1]{Algo.~\ref{#1}}

\newcommand{\specialcell}[2][c]{\begin{tabular}[#1]{@{}c@{}}#2\end{tabular}}
\newcommand{\note}[1]{\textcolor{red}{#1}}

\newcommand{\proj}{\textsc{Crescent}\xspace}
\newcommand{\mode}[1]{\underline{\textsc{#1}}\xspace}
\newcommand{\sys}[1]{\underline{\textsc{#1}}}

\newcommand{\no}[1]{#1}
\renewcommand{\no}[1]{}
\newcommand{\RNum}[1]{\uppercase\expandafter{\romannumeral #1\relax}}

\def\cA{{\mathcal{A}}}
\def\cF{{\mathcal{F}}}
\def\cN{{\mathcal{N}}}
\def\sc{{\mathsf{C}}}
\def\bh{{\mathbf{h}}}
\def\bp{{\mathbf{p}}}
\def\bx{{\mathbf{x}}}
\def\llhh{{\mathsf{LH}}}
\def\hhll{{\mathsf{HL}}}


%% file: sec_intro.tex
\section{Introduction}
\label{sec:intro}

Virtual Reality (VR) has the potential of becoming the next ubiquitous computing platform, after PCs and smartphones, revolutionizing a wide variety of domains such as healthcare~\cite{arvrhealthcare}, education~\cite{akccayir2017advantages}, remote communication~\cite{piumsomboon2018mini, oda2015virtual}, professional training~\cite{han2022augmented}, and industrial design~\cite{fiorentino2002spacedesign}.

Continued progress in the burgeoning field of VR depends critically on an efficient computing substrate, driven by the ever-growing requirement of immersive user experience and the miniaturization of device form factors.
DRAM communication energy is known to contribute significantly to the system energy consumption.
Recent studies show that DRAM energy alone can consume upward of 30\% of the total system energy consumption during VR video rendering~\cite{zhao2020deja, haj2021burstlink}.
The DRAM bottleneck will only become worse in the future with users questing for higher resolution and frame rate.

An effective approach to reduce DRAM traffic is framebuffer compression, which is universally implemented in modern mobile SoCs for compressing any traffic in and out of the DRAM.
A classic example is the Arm Frame Buffer Compressions (AFBC) technology, which is now in almost all of Arm's GPU, Video Codec, and Display Controller IPs~\cite{afbc}.

\paragraph{Idea.} Today's framebuffer compression algorithm is numerically lossless.
Being numerically lossless is, however, unnecessary to preserve perceptual fidelity: more compression opportunities arise when we turn our attention to \textit{perceptual lossless}.
Long-established psychophysical studies show that humans cannot discriminate colors that are close to each other~\cite{wright1934hue, krauskopf1992color}. Informally, this means that many colors, while differing in RGB values, are perceptually indistinguishable and thus can be encoded together --- a previously under-exploited opportunity for real-time image encoding.

Critically, the discrimination ability becomes even weaker (i.e., more colors are indistinguishable) in our \textit{peripheral} vision as objects move away from fixation~\cite{duinkharjav2022color,cohen2020limits,hansen2008color}.
The eccentricity-dependent weakening of color discrimination provides further opportunities for DRAM traffic compression:
VR displays, to provide immersive experiences, have a wide Field-of-View (FoV) of about 100\textdegree;
above 90\% of a frame's pixels are in the peripheral vision (outside 20 \textdegree)~\cite{albert2017latency, patney2016towards}.

\paragraph{Design.} Leveraging the unique color discrimination (in)ability of human visual system, we propose a new image compression algorithm for immersive VR systems.
We precisely formulate the color perception-aware encoding as a constraint optimization problem.
The formulation is non-convex and requires iterative solvers that are not amenable to real-time execution.
Leveraging empirical observations of human color discrimination abilities, we introduce a set of principled relaxations, which transform the compression problem into a convex optimization with an analytical solution.

The analytical solution, while avoiding iterative solvers, is still compute intensive and slow to execute in real-time.
Implemented as a GPU shader executing on the Adreno 650 GPU in Oculus Quest 2, a widely used mobile VR headset, the compression algorithm runs in a mere 2 FPS.
We propose lightweight hardware extensions for our encoding and decoding algorithms.
The new hardware exploits the inherent task-level and pipeline-level parallelisms in the algorithms and can be readily combined with existing Base-Delta (BD) encoding without changing the decoding hardware at all.

\paragraph{Results.}
We implement our architectural extensions in  RTL and synthesize the design using a TSMC 7 nm process node.
The compression algorithm reduces the memory traffic by 66.9\% compared to uncompressed images and by up to 20.4\% compared to the state-of-the-art real-time framebuffer compression~\cite{zhang2019distilling}.
We conduct IRB approved human subject study on 11 participants. Results suggest that our compression algorithm brings little visible artifacts to users.
In summary, this paper makes the following contributions:

\begin{itemize}
	\item We propose an image encoding scheme to reduce DRAM traffic in mobile VR systems. The scheme leverages the eccentricity-dependent color discrimination (in)ability of human visual systems.
	\item We show that the new encoding scheme can be formulated as a convex optimization problem with an analytical solution.
	\item We propose lightweight and modular hardware support to enable real-time encoding.
	\item ASIC synthesis and human subject studies show that the new encoding scheme reduces the DRAM traffic by 66.9\% with little to no subjective perceptual quality degradation.
\end{itemize}

The rest of the paper is organized as follows. \Sect{sec:bck} introduces the background. \Sect{sec:algo} describes our key compression algorithm. \Sect{sec:hw} introduces the co-designed hardware architecture. \Sect{sec:exp} discusses the experimental methodology, followed by the evaluation results in \Sect{sec:eval}. We relate our work to prior art in \Sect{sec:related} and conclude in \Sect{sec:conc}.

%% file: sec_brw.tex
\section{Background and Motivation}
\label{sec:bck}

We first introduce the background of human color perception and its eccentricity dependence, which form the psychophysical basis for our compress algorithm (\Sect{sec:bck:color}).
We then describe today's real-time frame compression algorithm, which forms an important baseline for our algorithm (\Sect{sec:bck:comp}).

\subsection{Eccentricity-Dependent Color Perception}
\label{sec:bck:color}

\paragraph{Colors and Color Spaces.} In a typical rendering pipeline, a color is usually described in the linear RGB space with three channels;
each channel is a floating point number between 0 and 1.
For output encoding, each channel in the linear RGB color space is transformed to the common sRGB color space, where each channel is an 8-bit integer between 0 and 255. This transformation is non-linear, called gamma encoding, and is described by the following function $f_{s2r}$, where $x \in [0, 1]$ represents a linear RGB channel value~\cite{srggamma, giorgianni2017color}:
\begin{align}
  f_{s2r}(x) := \begin{cases}
  \lfloor 12.92x \rfloor  & x \leq 0.0031308 \\
  \lfloor 1.055 x^{1/2.4} - 0.055 \rfloor & x > 0.0031308
  \end{cases}
  \label{eq:srgb}
\end{align}
Psychophysical studies on color discrimination commonly operate in the DKL color space~\cite{derrington1984chromatic, hansen2008color}, mainly because the DKL space models the opponent process in the human visual system. The DKL space is a linear transformation away from the linear RGB color space:
%
\begin{align}
    [R, G, B]^T = &\text{M}_{\text{RGB2DKL}} [K_1, K_2, K_3]^T
\end{align}
\noindent where $[R, G, B]$ is the color in the linear RGB space, $[K_1, K_2, K_3]$ is the color in the DKL space, and $\text{M}_{\text{RGB2DKL}}$ is a $3\times 3$ constant matrix (with the same coefficients, \begin{math}[[0.14, 0.17, 0.00], [-0.21, \\ -0.71, -0.07], [0.21, 0.72, 0.07]]\end{math}, as in Duinkharjav et al.~\mbox{\cite{duinkharjav2022color}}).

\begin{figure}[h]
    \centering
    \includegraphics[width=\columnwidth]{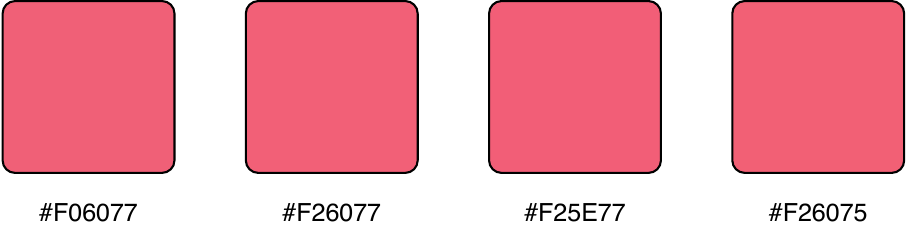}
    \vspace{-15pt}
    \caption{Human visual system can not discriminate colors that close to each other. These four colors differ in tristimulus values, but appear to be the same color.}
    \label{fig:colordisc}
\end{figure}

\paragraph{Color Discrimination.}
It is well-established that humans can not discriminate between colors that are close to each other~\cite{wright1934hue, krauskopf1992color}.
For instance, \Fig{fig:colordisc} shows four colors that have different sRGB values but appear to be the same.

\begin{figure*}[t]
    \centering
    \includegraphics[trim={30pt 0 0 0},clip,width=2\columnwidth]{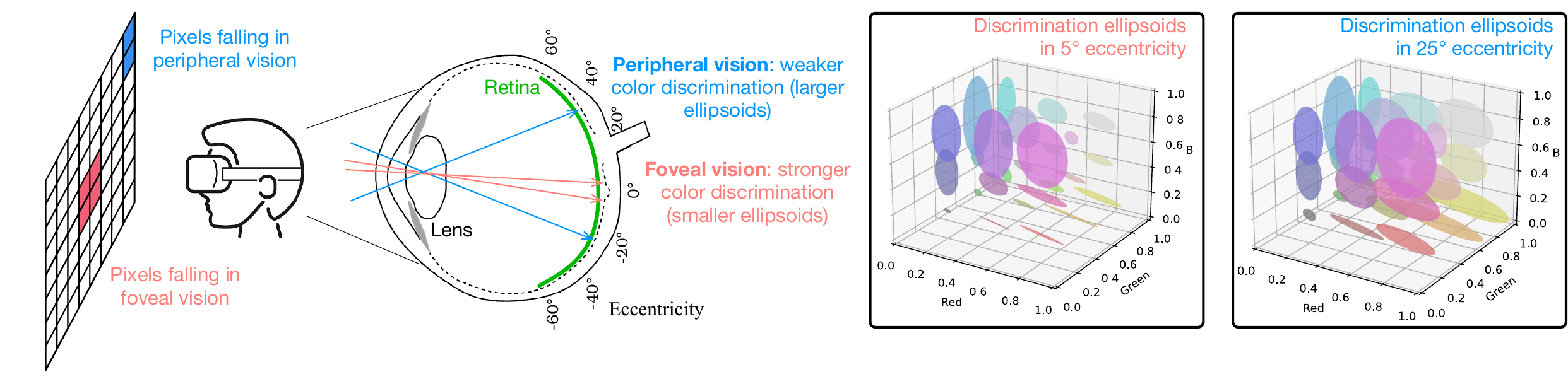}
    \caption{Color discrimination is eccentricity dependent. The discriminative ability is weaker as the eccentricity increases. As a result, the sizes of the discrimination ellipsoids increase with the eccentricity. The two plots on the right show the discrimination ellipsoids under a 5\textdegree~and a 25\textdegree~eccentricity, respectively, in the linear RGB color space (i.e., sRGB normalized to [0, 1] without gamma~\cite{srggamma, giorgianni2017color}). The discrimination ellipsoids in each plot are shown for 27 colors uniformly sampled in the linear RGB color space between [0.2, 0.2, 0.2] and [0.8, 0.8, 0.8].}
    \label{fig:eccdisc}
\end{figure*}

More formally, given a reference color $\mathbf{\kappa}$, there exists a \textit{set} of colors $\mathcal{E}_\mathbf{\kappa}$, in which all the colors are perceptually indistinguishable from $\mathbf{\kappa}$.
In a linear color space such as DKL and RGB, the set of equi-appearance colors in $\mathcal{E}_\mathbf{\kappa}$ form an \textit{ellipsoid}, whose center is $\mathbf{\kappa}$~\cite{macadam1942visual}. In the literature, such an ellipsoid is called a \textit{discrimination ellipsoid}~\cite{wyszecki2000color}.

\paragraph{Eccentricity Dependence.} Critically, human color discrimination ability is weaker in the peripheral vision~\cite{cohen2020limits, duinkharjav2022color}. That is, for a color $\mathbf{\kappa}$, its discrimination ellipsoid $\mathcal{E}_\mathbf{\kappa}$ is larger, i.e., includes more indistinguishable colors, as $\mathbf{\kappa}$ moves away from one's fixation.
\Fig{fig:eccdisc} shows two figures that plot the discrimination ellipsoids under a 5\textdegree~and a 25\textdegree~eccentricity, respectively, in the linear RGB color space.
Eccentricity is the angle from the center of the retina, a.k.a., current fixation or ``fovea''.
The ellipsoids in the 25\textdegree~plot are larger than those in the 5\textdegree~plot, suggesting that the color discrimination ability is weaker in peripheral vision.

Color discrimination becomes weaker in the visual periphery for three reasons.
First, the receptive field (RF) sizes of Retinal Ganglion Cells (RGCs) increase with eccentricity, a result of larger dendritic fields~\cite{rodieck1985parasol, dacey1993mosaic} and sparser RGC density in periphery~\cite{curcio1990topography}.
A large RF means that a RGC integrates signals from a larger spatial area, leading to more blurring in the (spatial) frequency domain.
Second, cone cells (which are photoreceptors responsible for vision under normal daylight) become larger in size as eccentricity increases~\cite{curcio1990human}, also contributing blurring in spatial frequency.
Finally, the distribution of cone cells on our retina is extremely non-uniform: over 95\% of the cone cells are located in the central region of the retina (i.e., fovea) with an eccentricity of below 5 \textdegree~\cite{curcio1990human, Song:2011:ConeDensity}.
The density of the cone cells decreases drastically in the visual periphery, which is, thus, significantly under-sampled spatially.

The full color discrimination function $\Phi$, expressed below, is thus parameterized by both the reference color $\mathbf{\kappa}$ and the eccentricity $\mathbf{e}$:
\begin{align}
	\Phi:  (\mathbf{\kappa}, \mathbf{e}) \mapsto (a, b, c)
\end{align}
\noindent where $(a, b, c)$ represents the semi-axes lengths of the discrimination ellipsoid belonging to color $\mathbf{\kappa}$ at an eccentricity $\mathbf{e}$ in the DKL color space~\cite{derrington1984chromatic}, a common color space for color perception experiments. Given $(a, b, c)$, $\mathcal{E}_\mathbf{\kappa}$, the discrimination ellipsoid of color $\kappa$ in the DKL space, is given by:
\begin{align}
	\frac{(x-\kappa_1)^2}{a^2} + \frac{(y-\kappa_2)^2}{b^2} + \frac{(z-\kappa_3)^2}{c^2} = 1
\end{align}
\noindent where $(\kappa_1, \kappa_2, \kappa_3)$ represent the three channels of the color $\kappa$.

The function $\Phi$ can be implemented using a Radial Basis Function (RBF) network~\cite{duinkharjav2022color}, which is extremely efficient to implement on GPUs in real time. In our measurement on Oculus Quest 2 VR headset using Oculus' OVR Metrics Tool~\cite{ovrtool}, evaluating RBF network runs in 72 FPS, matching the display refresh rate while consuming sub 1~mW power.

AR and VR headsets, in providing an immersive experience, usually have a wide FoV that is above 100\textdegree. Therefore, the vast majority of the pixel colors will fall in the peripheral vision. The eccentricity-dependent color discrimination (in)abilities of human visual system gives opportunities to better image compression that this paper exploits.

\subsection{Real-Time Frame Compression}
\label{sec:bck:comp}

\begin{figure}[t]
    \centering
    \includegraphics[width=\columnwidth]{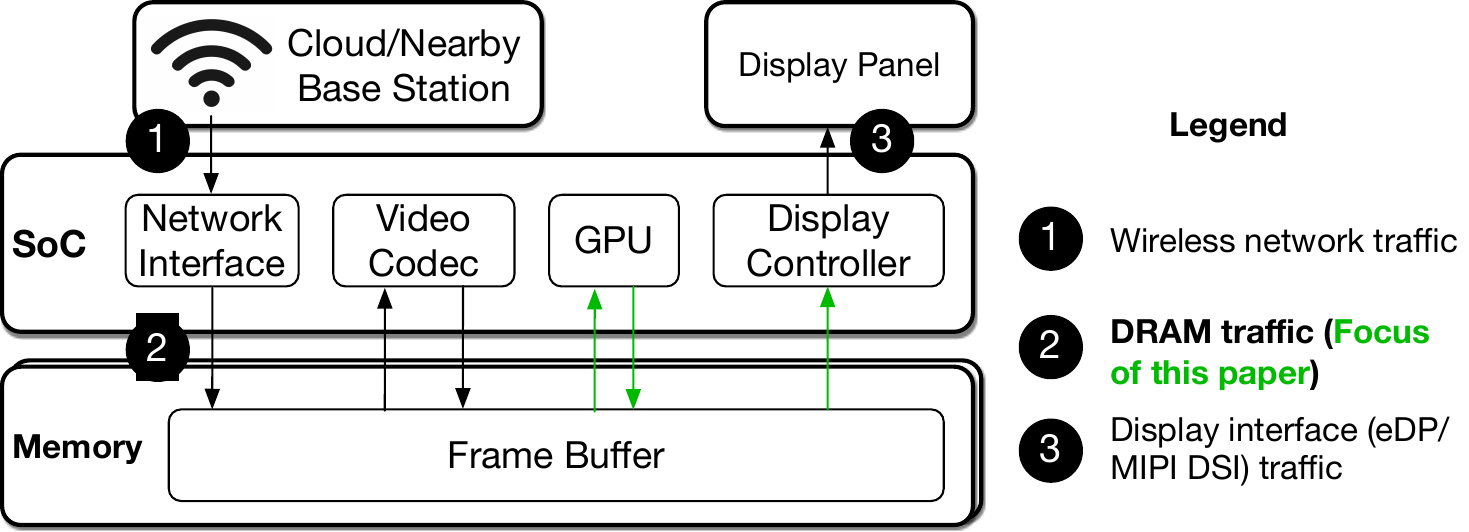}
    \caption{Different types of data communication traffic in a VR system. This paper focuses on reducing DRAM traffic.
    }
    \label{fig:traffic}
\end{figure}

\paragraph{DRAM Traffic.} A variety of data communication traffics occur on a VR system, as illustrated in \Fig{fig:traffic}, such as the traffic through DRAM, the display interface, and the wireless communications with a remote rendering server.
This paper focuses on reducing the DRAM traffic, which occurs when the different Intellectual Property (IP) blocks in the SoC communicate with each other during rendering.

Each frame, the GPU writes the frame data to the frame buffer in the GPU, which are then read by the display controller.
It is these DRAM traffics (i.e., GPU $\leftrightarrow$ frame buffer $\leftrightarrow$ DRAM controller) that this paper focuses on reducing.
When rendering a VR (360\textdegree) video, additional DRAM traffics occur between the network interface controller, the video codec, and the GPU~\cite{leng2019energy}. While not explicitly targeted in this paper, these traffics can also potentially be reduced by our compression algorithm, especially in scenarios where remotely rendered frames are transmitted one by one (rather than as a video)~\mbox{\cite{Krajancich:2020:spatiotemp_model, kaplanyan2019deepfovea}}.



Reducing DRAM traffic is critical. It is well-established that data transfer and memory access energy is known to far out-weigh the energy consumption of computation. For instance, compared to a Multiple-Accumulate (MAC) operation on 1-Byte fixed-point data, transferring one Byte of information through DRAM consumes 800 $\times$~\cite{liu2019intelligent, liu2022augmented} higher energy.
Reducing DRAM traffic in a visual computing system has been a main research focus in recent years~\cite{kodukula2021rhythmic, zhang2019distilling, haj2021burstlink}.

\paragraph{Framebuffer Compression Algorithms.}
An effective and commonly used approach to reduce DRAM traffic in a rendering system is framebuffer compression, which compresses and uncompresses every frame in and out of the DRAM.
To ensure a low \textit{per-frame} latency, compression in VR must be done on a per-frame basis, precluding video compression methods such as H.265/VP5, which necessarily require buffering a sequence of frames before compression~\cite{poynton2012digital, richardson2011h}.
Offline image compression methods such as JPEG and PNG are rarely used in framebuffer compression as they are too compute-intensive. For instance, JPEG requires chroma subsampling, transforming images to a frequency space followed by quantization and Huffman encoding~\cite{wallace1991jpeg}.

\begin{figure}[t]
    \centering
    \includegraphics[width=\columnwidth]{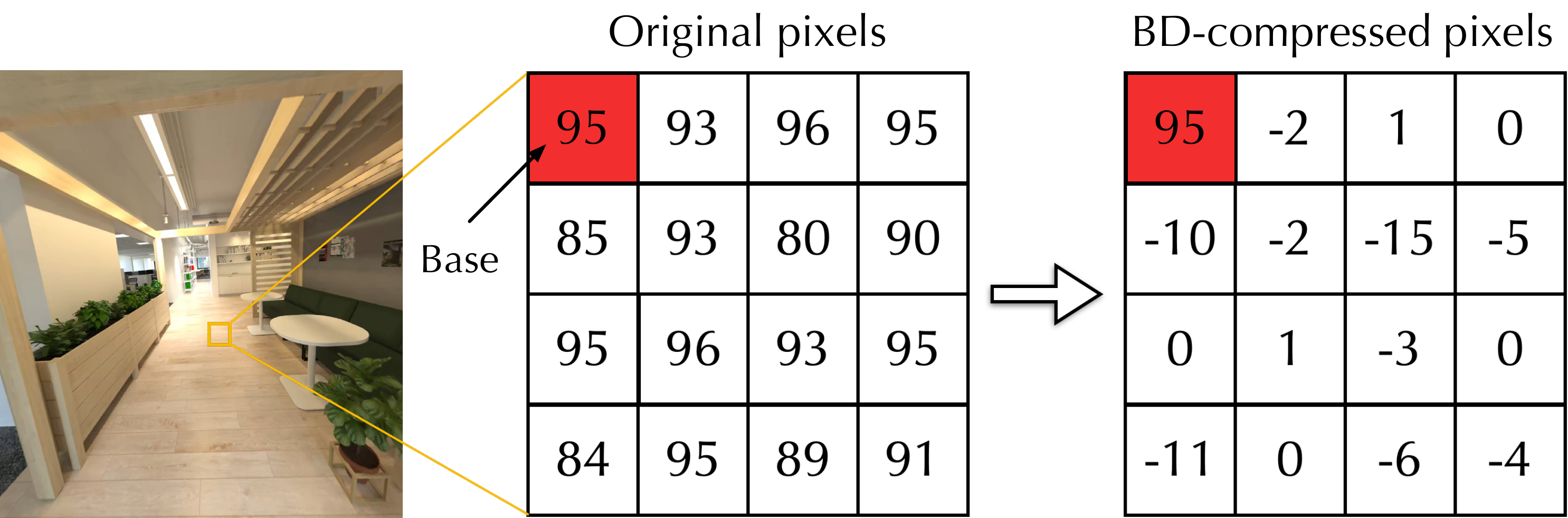}
    \caption{Base + Delta (BD) compression, which works in the sRGB color space. For each pixel tile (4$\times$4 here), we find a base pixel (95 here), and calculate the $\Delta$ of all other pixels from the base pixel. The $\Delta$ are smaller in magnitude and thus require fewer bits to encode. The same compression strategy is applied to all three color channels.}
    \label{fig:bd}
\end{figure}

Today's framebuffer compression methods universally use a much faster base+delta (BD) strategy. \Fig{fig:bd} uses a simple example to illustrate the basic idea behind BD, which compresses each color channel and each pixel tile individually.
The tile size in \Fig{fig:bd} is 4$\times$4. In each tile, BD chooses a base pixel and then calculates the $\Delta$s/offsets between all other pixels and the base pixel. In the example of \Fig{fig:bd}, the base pixel is the first pixel. The $\Delta$s will necessarily have smaller magnitudes compared to the original pixel values and, thus, require fewer bits to encode.

The BD compression algorithm is lightweight: it works completely in the image space, as opposed to the frequency domain which requires an additional, compute-intensive transformation (e.g. Fast Fourier Transform or Discrete Cosine Transformation); it requires only fixed-point addition arithmetics; it is also embarrassingly parallel.
Therefore, the basic BD strategy is universally implemented in today's mobile SoCs for compressing any traffic in and out of the DRAM. A classic example is the Arm Frame Buffer Compressions (AFBC) technology, which is now in almost all of Arm's GPU, Video Codec, and Display Controller IPs~\cite{afbc}.

%% file: sec_algo.tex
\section{Color Perception-Aware Compression}
\label{sec:algo}

This section introduces a color perception-aware image encoding and decoding algorithm.
We start by describing the high-level ideas (\Sect{sec:algo:ideas}), followed by a precise problem formulation in the form of constraint optimization (\Sect{sec:algo:formulation}).
We then show how this optimization problem has an analytical solution when relaxed to a convex problem (\Sect{sec:algo:sol}).
We then describe the full compression algorithm (\Sect{sec:algo:algo}).

\subsection{Key Ideas}
\label{sec:algo:ideas}

The basic BD algorithm is numerically lossless.
Our observation is that numerically lossless compression is unnecessary to preserve perceptual equivalence --- because of the inherent the color discrimination (in)ability of human visual system.

\paragraph{Intuition.}
The basic BD algorithm encodes all the $\Delta$s in a tile (off of a base pixel) rather than the original pixel values.
Thus, to improve the compression ratio over BD we must reduce the magnitude of the $\Delta$s, which, intuitively, requires bringing pixels \textit{more similar} to each other.

Under a numerically lossless constraint, however, the $\Delta$s between pixels are fixed.
Our idea is to relax the constraint from numerical lossless to \textit{perceptually lossless}.
In this way, we could adjust pixel color values, as long as each pixel color does not go beyond its discrimination ellipsoid, to minimize the total number of bits required to encode the $\Delta$s.
This encoding is numerically lossy as we intentionally change the color values, but will preserve the perceptual quality.

\begin{figure}[h]
    \centering
    \includegraphics[width=\columnwidth]{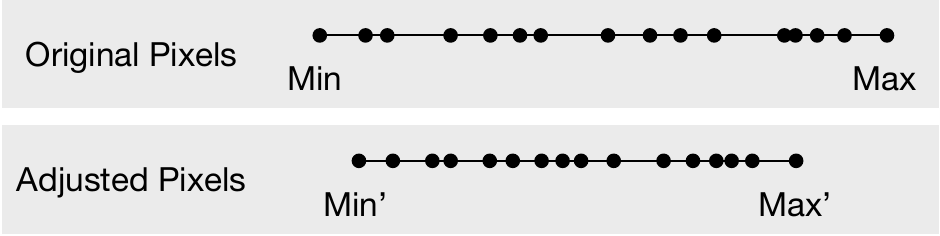}
    \caption{An intuition illustration of our perceptual-aware compression, where pixel values are adjusted to be more similar to each other by leveraging the inherent human color discrimination thresholds.}
    \vspace{-5pt}
    \label{fig:intuition}
\end{figure}

\paragraph{An Example.} More concretely, consider the example in \Fig{fig:intuition}, which shows 16 pixels in a tile on an axis.
The number of bits required to encode the entire tile is (ignoring any metadata for now):
\begin{align}
	B = B_0 + N \times B_D \\
	B_0 = 8, N = 15, B_D = \lfloor log_2(Max - Min + 1)  \rfloor
\end{align}
\noindent where $B_0$ being 8 denotes that we need 8 bits to encode a base pixel (assuming the common 8-bit per-channel encoding), and $N$ being 15 denotes that there are 15 other pixels.
$B_D$ denotes the number of bits required to encode the $\Delta$ of each of the 15 non-base pixels.

The minimum value of $B_D$ occurs when the base pixel is chosen to be within [$Min$, $Max$], in which case $B_D = \lfloor log_2(Max - Min + 1)  \rfloor$.
This is because the number of bits to encode each $\Delta$ must be the same\footnote{It is possible, but uncommon, to vary the number of bits to encode the $\Delta$s in a tile with more hardware overhead. Following prior work~\cite{zhang2019distilling}, this paper assumes that one single bit-length is used to encode all $\Delta$s in a tile. We consider variable bit-length an orthogonal idea to this paper.}, so we must accommodate the \textit{largest possible} $\Delta$, which is the difference between the maximum and minimum pixels in the tile.
Therefore, to improve compression ratio we must reduce $(Max-Min)$.

The bottom example in \Fig{fig:intuition} illustrates what would happen when we relax the compression constraint to be perceptually lossless.
The adjusted pixel values deviate from the original values, but as long as they still within the respective ellipsoids, $(Max-Min)$ is reduced without affecting perceptual quality.

It is worth noting that to obtain the highest compression rate it is necessary to adjust interior pixels, as is the case in this example. The central challenge we address in this paper is how to design a principled algorithm that maximizes the bit reduction while being lightweight to execute in real time.

%

\subsection{Problem Formulation}
\label{sec:algo:formulation}

Our compression algorithm works on top of the baseline BD algorithm. Our goal is to adjust pixel colors to minimize the bit-length required to encode the $\Delta$s in a tile.
The adjusted pixel tile then goes through any existing BD compression method.
Critically, color adjustment must not violate the perceptual constraints.
Therefore, we formulate our compression as a constraint optimization problem:
%
%
\begin{subequations}
\begin{align}
  \underset{\bp}{\text{argmin}}~& \sum_{\sc\in\{R,G,B\}}log_2 \lfloor max\{ f_{s2r}(\bp^{\sc}) \} - min\{ f_{s2r}(\bp^{\sc}) \} + 1\rfloor,\label{eq:opt:obj} \\
  \text{where}~& \mathbf{p} := [p_0,~ p_1,~ \cdots,~ p_{N-1}], \label{eq:opt:var} \\
  & \bp^{\sc} := [p_0^{\sc},~ p_1^{\sc},~ \cdots,~ p_{N-1}^{\sc}],~~~\sc\in\{R,G,B\} \label{eq:opt:pixel} \\
  s.t. ~~~ & \forall p_i \in \mathbf{p} ~~~ p_i \in \mathcal{E}_{p_i} \label{eq:opt:con}
\end{align}
\label{eq:opt}
\end{subequations}
\noindent where $\bp$ is the optimization variable, which is the collection of $N$ pixels in a tile (\Equ{eq:opt:var}); $p_i^\sc$ denotes channel $\sc$ (R, G, or B) of $i$-th pixel in the linear RGB space.

The constraints (\Equ{eq:opt:con}) provide the (convex) ellipsoid boundary for each pixel to move while maintaining perception quality.
$f_{s2r}(\cdot)$ is the non-linear transformation from RGB to sRGB space (\Sect{sec:bck:color}), which is ultimately where bit encoding takes place.
The objective function (\Equ{eq:opt:obj}) minimizes the bit cost for encoding the $\Delta$s across all channels (it is a constant cost to encode the base pixel, e.g., 8 in the common sRGB encoding).
This optimization formulation is applied to each pixel tile independently.


Unfortunately, this optimization problem is impractical to be solved in real-time, because the objective function is non-convex due to the non-linearity of min, max, floor, and $f_{s2r}(\cdot)$.
Empirically, we also find that the popular solvers in Matlab spend hours while still being stuck in local optima.


\paragraph{Relaxation.}
We introduce two relaxations that turn the problem into a convex optimization.
Critically, while general convex optimization requires iterative solvers (e.g., gradient descent or Newton's method~\cite{boyd2004convex}), our relaxed problem is one such that it has an analytical solution.
The relaxations keep the same constraints as before (\Equ{eq:opt:con}) and, thus, still enforce the perceptual quality.

The first relaxation is based on the empirical observation that most discrimination ellipsoids are elongated along the either the Red or the Blue axis.
See the discrimination ellipsoids in \Fig{fig:eccdisc} for an illustration.
This makes sense as human visual perception is most sensitive to green lights~\cite{sharpe2005luminous, wyszecki2000color} and, thus, has the least ``wiggle room'' along the Green axis.



Our idea thus is to, instead of minimizing the bit costs across \textit{all} three axes, minimize along \textit{only} the Red or the Blue axis (while still having the flexibility of adjusting all the channels of all the pixels in a tile).
Using the Blue axis an example, this relaxation yields following new objective function in \Equ{eq:opt:obj-r1}:
\begin{subequations}
  \begin{align}
    \underset{\bp}{\text{argmin}}~& log_2 \lfloor max\{ f_{s2r}(\bp^{B}) \} - min\{ f_{s2r}(\bp^{B}) \} + 1\rfloor,\label{eq:opt:obj-r1} \\
    \Rightarrow \underset{\bp}{\text{argmin}}~& max\{ f_{s2r}(\bp^{B}) \} - min\{ f_{s2r}(\bp^{B}) \},\label{eq:opt:obj-r2} \\
    \xRightarrow{\sim} \underset{\bp}{\text{argmin}}~& max\{ \bp^{B} \} - min\{ \bp^{B} \}.\label{eq:opt:obj-r3}
  \end{align}
  \label{eq:opt-sim}
\end{subequations}
%
%

Second, the objective function in \Equ{eq:opt:obj-r1} can be transformed to \Equ{eq:opt:obj-r2} without sacrificing solution optimality, because $log_2\lfloor \cdot \rfloor$ is monotonically non-decreasing.
We then remove the non-linear RGB to sRGB transformation function $f_{s2r}(\cdot)$.
This removal does not preserve the solution optimality, but gives us a convex objective function in \Equ{eq:opt:obj-r3}.

\paragraph{Proof of Convexity.} Let the objective function $max\{ \bx \} - min\{ \bx \}$ be $g(\bx): \mathbb{R}^N \rightarrow \mathbb{R}$. To prove $g(\bx)$ is convex, we must show: 
$\forall \bx_1, \bx_2 \in \mathbb{R}^N$ and $t \in [0, 1]$, $g(t \bx_1 + (1-t) \bx_2) \leq t g(\bx_1) + (1-t) g(\bx_2)$.

\begin{proof}
Observe that: $g(t \bx_1 + (1-t) \bx_2) := max(t \bx_1 + (1-t) \bx_2) - min(t \bx_1 + (1-t) \bx_2)$.

We know $max(t \bx_1 + (1-t) \bx_2) \leq max(t \bx_1) + max((1-t) \bx_2) = t~max(\bx_1) + (1-t)~max(\bx_2)$. 
Similarly we can derive: $min(t \bx_1 + (1-t) \bx_2) \geq t~min(\bx_1) + (1-t)~min(\bx_2)$.

Therefore, $g(t \bx_1 + (1-t) \bx_2) \leq (t~max(\bx_1) + (1-t)~max(\bx_2)) - (t~min(\bx_1) + (1-t)~min(\bx_2)) = t g(\bx_1) + (1-t) g(\bx_2)$.
\end{proof}

\subsection{Analytical Solution Intuition}
\label{sec:algo:sol}

\begin{figure}[t]
  \centering
  \subfloat[Case 1: $HL>LH$.]{
        \label{fig:ncbp}
        \includegraphics[width=\columnwidth]{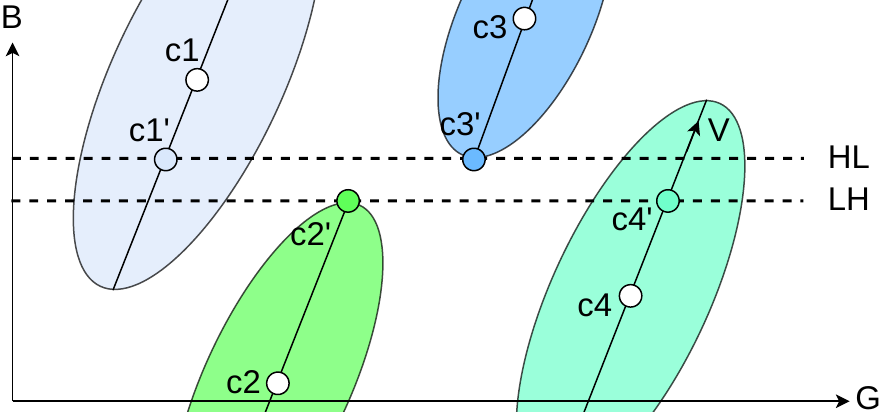}
  }\\
  \subfloat[Case 2: $HL<LH$.]{
        \label{fig:cbp}
        \includegraphics[width=\columnwidth]{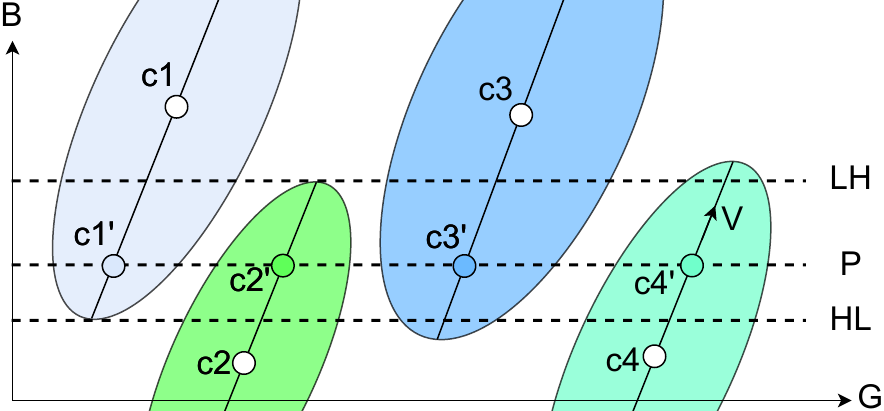}
  }
  \caption{The two cases in adjusting color values to minimize the $\Delta$ along the Blue axis. For simplicity, we draw the ellipsoids in the B-G plane. The empty markers ($C_0, C_1, C_2, C_3$) denote the original colors and the solid markers ($C_0', C_1', C_2', C_3'$) denote the adjusted colors.
  In both cases, colors are adjusted along the extrema vector $\mathbf{V}$.
   }
  \label{fig:movc}
\end{figure}

\begin{figure*}
    \centering
    \includegraphics[width=\linewidth]{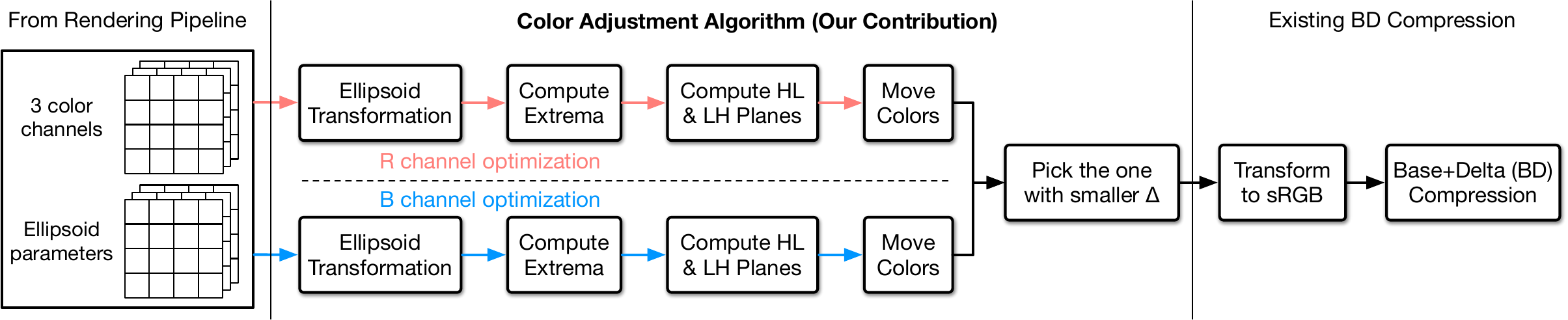}
    \caption{Overview of our algorithm and how it fits in existing rendering and compression pipeline. Our algorithm takes a tile of pixels and their corresponding discrimination ellipsoid parameters, and generate an adjusted pixel tile, which then goes through existing BD encoding.}
    \label{fig:algo}
\end{figure*}

The relaxations introduced before lead to an analytical solution without requiring iterative solvers.
Observe that the objective function in \Equ{eq:opt:obj-r3} minimizes the difference between the maximum and minimum values along the Blue axis.
To achieve that, the intuition is that we must move the colors closer to each other along the Blue axis while making sure the adjusted colors stay within the respective discriminative ellipsoids.

Exactly how to move the colors falls into two cases.
\Fig{fig:movc} illustrates the two cases using two examples.
Without losing generality, we choose to optimize along the Blue axis in these examples (the case along the Red axis is in principle the same), and we plot the projection of the ellipsoids onto the B-G plane for better visualization.

In the first case (\Fig{fig:ncbp}), there is no single plane that cuts across all ellipsoids.
This is because the Lowest of the Highest points of all ellipsoids ($\llhh$) is lower than the Highest of the Lowest points of all ellipsoids ($\llhh$).
The optimal strategy is to move all the colors higher than $\hhll$ toward $\hhll$ and move all the colors lower than $\llhh$ toward $\llhh$.
The movement is necessarily executed along the \textit{extrema vector}, which is the vector that connects the highest and the lowest point of an ellipsoid.
After the adjustment, the Blue channels across all the pixels are either $\hhll$ or $\llhh$.
That is, the maximum $\Delta$ along the Blue axis is now $\hhll - \llhh$, which is the smallest gap we can get the Blue channels to be without going outside the ellipsoid boundaries.

In the second case (\Fig{fig:cbp}), there is a common plane (\fixme{P}) that cuts across all four ellipsoids. In fact, there are infinitely many such planes, because $\llhh$ is higher $\hhll$; thus, any plane between $\llhh$ and $\hhll$ will cut across all ellipsoids.
In this case, we can simply pick any such plane and move all the colors to that plane.
For the simplicity of implementation, we choose the average of the $\llhh$ and the $\hhll$ planes as the common plane and move colors along the extrema vectors.
In this way, the Blue channel value is exactly the same for all pixels, requiring no $\Delta$ bit for the Blue channel.

\subsection{Overall Compression Algorithm}
\label{sec:algo:algo}

We illustrate how our color adjustment algorithm fits in the overall rendering and compression pipeline in \Fig{fig:algo}.
Our adjustment algorithm takes as inputs a tile of pixels (each with three channels) and the parameters of their corresponding discrimination ellipsoids.
The algorithm generates the perceptually-adjusted pixel tile as the output.
We apply the same color adjustment strategy along both the Blue and the Red axis for each tile, and pick the better one in the end.

It is worth noting that our algorithm does not directly perform compression in itself; it simply adjusts pixel colors so that the (numerically lossless) BD encoding later can achieve higher compression rate.
Specifically, the adjusted pixel tile will be first transformed from the linear RGB to the sRGB space, which then goes through the usual BD compression.

\paragraph{Ellipsoid Transformation.}
The first step in our algorithm is to transform the discrimination ellipsoids from the DKL space to the linear RGB space, which is where color adjustment takes place (\Sect{sec:algo:sol}).
While ellipsoids are axis-aligned in the DKL color space~\cite{duinkharjav2022color}, they will not be axis-aligned after the linear transformation from the DKL to the RGB color space.
Therefore, an ellipsoid in the linear RGB space has to take the form of a general quadric surface:
\small
\begin{align}
	Ax^2+By^2+Cz^2+Dx+Ey+Fz+Gxy+Hyz+Izx+1 = 0
	\label{eq:ellipsq}
\end{align}
\normalsize

Transforming an axis-aligned ellipsoid in the DKL space to an ellipsoid in the linear RGB amounts to the following matrix multiplication:
%
\small
\newcommand{\bigzero}{\mathbf{0}}
\newcommand{\rvline}{\hspace*{-\arraycolsep}\vline\hspace*{-\arraycolsep}}
\newcommand{\rhline}{\vspace*{-\arraycolsep}\hline\vspace*{-\arraycolsep}}
\begin{align}
 	\begin{bmatrix}
 		A \\
 		B \\
 		C \\
 		D \\
 		E \\
 		F \\
 		G \\
 		H \\
 		I \\ 
 	\end{bmatrix} =
	\begin{bmatrix}
		(T\odot T)^\top & \vline & \bigzero \\
  \cmidrule(lr){1-3}
		\bigzero & \vline & T\\
  \cmidrule(lr){1-3}
   \begin{bmatrix}
2T_{00}T_{01} & 2T_{10}T_{11} & 2T_{20}T_{21} \\
2T_{01}T_{02} & 2T_{11}T_{12} & 2T_{21}T_{22} \\
2T_{00}T_{02} & 2T_{10}T_{12} & 2T_{20}T_{22} \\
  \end{bmatrix} & \vline & \bigzero\\
	\end{bmatrix} \times \nonumber 
	& \begin{bmatrix}
		1/a^2t \\
		1/b^2t \\
		1/c^2t \\
		-2\kappa_1 /a^2t \\
		-2\kappa_2 /b^2t \\
		-2\kappa_3 /c^2t 
	\end{bmatrix},\\
	t = 1-\big(\frac{\kappa_1^2}{a^2}+\frac{\kappa_2^2}{b^2}+\frac{\kappa_3^2}{c^2}\big)
\end{align}
\normalsize
\noindent where $T=\Big[ \begin{smallmatrix} T_{00} && T_{01} && T_{02} \\ T_{10} && T_{11} && T_{12} \\ T_{20} && T_{21} && T_{22} \end{smallmatrix} \Big]$ is the constant $\text{M}_{\text{RGB2DKL}}$ matrix in \Sect{sec:bck:color}, $\odot$ is element-wise product, $(\kappa_1, \kappa_2, \kappa_3)$ is the color in DKL space, and $(a, b, c)$ are the semi-axis lengths of $\kappa$'s discrimination ellipsoids. The derivation uses basic linear transformations and is omitted here due to space constraints.

\paragraph{Color Adjustment.}
Once we have the ellipsoids in the linear RGB space, we can perform color adjustment, which, as illustrated in \Fig{fig:movc} and described in \Sect{sec:algo:sol}, is done in three steps: 1) compute the extrema, i.e., the highest and the lowest point, of each ellipsoid; 2) compute $\llhh$ and $\hhll$ based on the extrema of all ellipsoids; 3) compare $\llhh$ and $\hhll$ and move colors along extrema vectors accordingly.
Step 2 and 3 are relatively straightforward, so here we focus on the mathematical details of Step 1.

Extrema along the Blue axis can be computed by taking the partial derivatives of the ellipsoid equation along the Red and Green axes:
%
\begin{subequations}
    \begin{align}
        \frac{dz}{dx} &= 2Ax+Gy+Iz+D = 0 \label{eq:der1} \\
        \frac{dz}{dy} &= Gx+2By+Hz+E = 0 \label{eq:der2}
    \end{align}
    \label{eq:der}
\end{subequations}
These partial derivatives give us two planes, the intersection of which is a vector $\mathbf{v}$ that connects the two extrema. The extreme vector $\mathbf{v}$ is calculated by taking the cross product of the normal vectors of the two planes:
\begin{align}
  \mathbf{v} = (2A,G,I) \times (G,2B,H)
  \label{eq:cross}
\end{align}
%
%
The two extrema points $H$ and $L$ are then calculated by finding the intersection of $\mathbf{v}$ and the ellipsoid:

\begin{subequations}
    \begin{align}
    \mathbf{x} &:= (x_1, x_2, x_3) = \text{M}_{\text{RGB2DKL}} \times \mathbf{v}^{T}  \label{eq:int_trans}\\
    t & = 1/\sqrt{\frac{x_1^2}{a^2}+\frac{x_2^2}{b^2}+\frac{x_3^2}{c^2}} \label{eq:int_fac} \\ 
    H & = \text{M}_{\text{RGB2DKL}}^{-1} \times (\kappa_1 + x_1 t, \kappa_2 + x_2 t, \kappa_3 + x_3 t)^{T} \nonumber \\
    L & = \text{M}_{\text{RGB2DKL}}^{-1} \times (\kappa_1 - x_1 t, \kappa_2 - x_2 t, \kappa_3 - x_3 t)^{T} \label{eq:int_pt}
    \end{align}
    \label{eq:int}
\end{subequations}

\noindent where $\kappa$ is the pixel color in the DKL space, $(a, b, c)$ are DKL ellipsoid parameters, and $\text{M}_{\text{RGB2DKL}}$ is the RGB to DKL transformation matrix (\Sect{sec:bck:color}). We omit the derivation details due to space constraints, but the derivation amounts to a simple application of line-ellipsoid intersection and linear transformations between RGB and DKL space.

\paragraph{Remarks on Decoding.}
One desired byproduct of our algorithm is that it requires \textit{no} change to the existing framebuffer decoding scheme --- our color adjustment algorithm simply changes the input to BD.
During decoding (e.g., by the display controller), the existing BD decoder will construct the sRGB values from the BD-encoded data, which are then sent to the display.
The exact BD encoding format varies across implementations and is not our focus.
We assume the encoding format described in Zhang et al.~\cite{zhang2019distilling}.

\begin{figure*}[t]
    \centering
    \includegraphics[width=\linewidth]{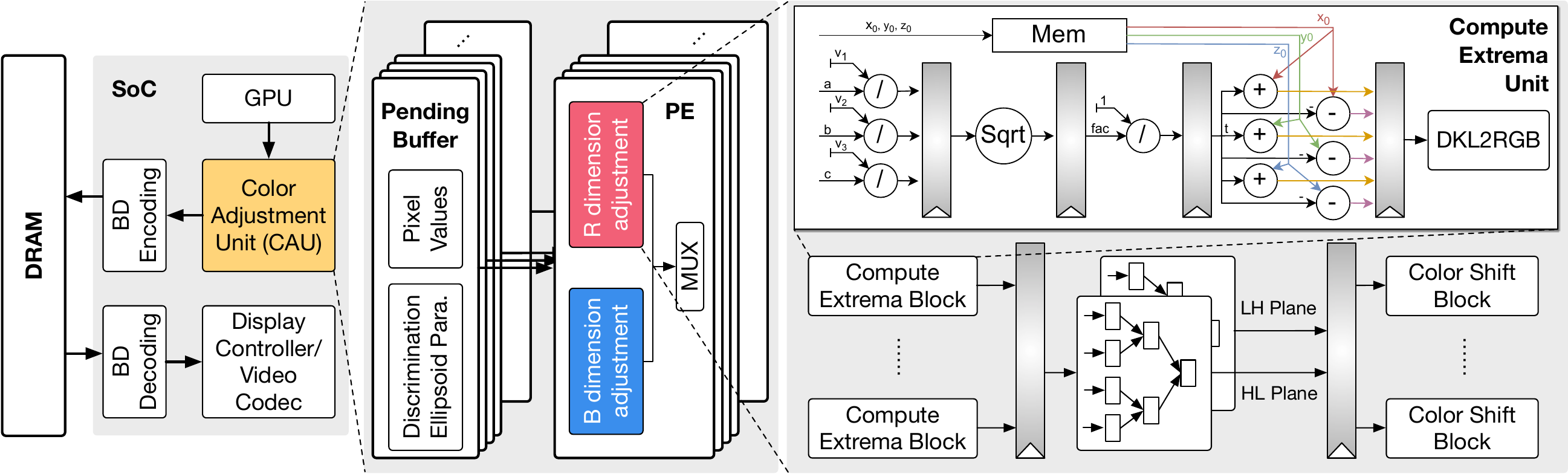}
    \caption{Illustration of the hardware support, which we dub Color Adjustment Unit (CAU) for our image encoding and how CAU interfaces with the rest of the SoC. Internally, the CAU uses an array of PEs, each of which adjust colors for one tile of pixels. CAU is fully pipelined to accept a new tile every cycle from the Pending Buffer, which receives the rendered pixels and their discrimination ellipsoids from the GPUs.}
    \label{fig:hw}
\end{figure*}

%% file: sec_hw.tex
\section{Hardware Architecture}
\label{sec:hw}

The analytical compression algorithm, while avoiding iterative solvers, is still compute intensive and slow to execute in real-time.
We implement it as a GPU shader executing on the Adreno 650 GPU in Oculus Quest 2, a widely used mobile VR headset.
The compression algorithm runs in a mere 2 FPS.
This section describes a lightweight hardware design that accelerates the compression algorithm.
\Sect{sec:hw:overview} describes how our custom hardware fits into the overall system and \Sect{sec:hw:cu} describes the hardware details.

\subsection{Hardware Overview}
\label{sec:hw:overview}



\Fig{fig:hw} provides an overview of our architectural extension, dubbed the Color Adjustment Unit (CAU), and how CAU fits into existing mobile SoCs.
The CAU executes the pixel adjustment algorithm described in \Sect{sec:algo}. The CAU reads its input from an on-chip buffer, which stores the pixels and the discrimination ellipsoid parameters generated by the GPU.
Following prior work~\cite{duinkharjav2022color}, we assume that the GPU is responsible for generating the per-pixel discrimination ellipsoids. The generation algorithm is a lightweight RBF network (\Sect{sec:bck:color}).
In our measurement, the ellipsoid generation algorithm on Oculus Quest 2 runs at the maximum display refresh rate (72 FPS) while consuming less than 1~mW measured using Oculus' OVR Metrics Tool~\cite{ovrtool}.

The output of the CAU enters the existing BD framebuffer encoder, which writes the encoded data to the DRAM.
Any frame read out from the DRAM, e,g., by the Displayer Controller IP block when sending the frame to the display, will enter the BD decoder, which reconstructs the sRGB pixels. The figure provides a visual confirmation that our algorithm 1) works on top of, rather than replaces, BD encoding, and 2) does not change the decoding architecture.


\subsection{Color Adjustment Unit}
\label{sec:hw:cu}

Internally, the CAU consists of an array of Processing Elements (PEs), each of which is designed to adjust colors for \textit{one tile} of pixels, which in our current design is assumed to be $4\times 4$.
Each PE interfaces with a dedicated Pending Buffer, which holds all the information of the pixel tiles generated from the GPU.
Having more PEs will allows the system to compressing multiples tiles simultaneously.


\paragraph{Pipelining.} The PE is fully pipelined to accept a new tile every cycle.
\Fig{fig:hw} illustrates the detailed architecture, which has three main phases, each of which is internally pipelined.
The first phase computes the extrema.
The next phases use reduction trees to calculate $\hhll$ and $\llhh$ from the extrema.
The final phase move the colors along the extrema vector.

\paragraph{Compute Extrema Blocks.}
This component calculates the extrema of all the pixels in a tile, which is naturally parallelizable across pixel and, thus, has multiple parallel units, each of which is responsible for one pixel.
The top-right box in \Fig{fig:hw} illustrates the microarchitecture.
This is the most compute intensive block in the CAU, since it involves multiple divisions and square root operations.
The division and square root hardware implements \Equ{eq:int_fac}, and the adder and subtractor circuit implements \Equ{eq:int_pt}.
The DKL-RGB transformations in \Equ{eq:int_pt} and \Equ {eq:int_trans} are implemented through matrix vector multiplication executed on a $3\times 3$ MAC array.


\paragraph{Compute Planes Blocks.}
The extrema calculated before enters this unit, which finds the channel value for the $\hhll$ plane (maximum of the minima) and $\llhh$ (minimum of the maxima) plane.
We implement this stage using two reduction (comparator) trees to generate both planes simultaneously. 

\paragraph{Color Shift Blocks.}
This block takes the original color values and the two planes as input and outputs the modified color values. This phase is control-flow heavy, as it involves multiple condition checks, e.g., testing the relationship between a point and a plane.
A custom datapath in CAU avoids much of the inefficiencies surrounding control flows that are detrimental to GPU performance.
This hardware is a relatively straightforward mapping from the algorithm.

\paragraph{Pending Buffer.}
The Pending Buffers store intermediate pixels and their discrimination ellipsoids from the GPU before they are consumed by the CAU.
Each buffer is interfaced with a dedicated PE and, thus, contains the data for all the pixel tiles to be consumed by the PE.

The buffers must be properly sized so as to not stall or starve the CAU pipeline.
In order to be independent of the exact GPU microarchitecture details, we make a conservative estimation of the buffer size.
In particular, we allocate enough space in the buffer such that it can hold all the pixels generated by the GPU in each CAU cycle \textit{even if} the GPU is fully utilized, in which case each shader core in the GPU generates 1 pixel/GPU cycle.
Note that the GPU and CAU cycle times need not be the same.
The number of PEs in a CAU must be properly decided so as to not stall either the GPU nor the CAU, as we will discuss in \Sect{sec:eval:oh}.



%% file: sec_es.tex
\section{Experimental Methodology}
\label{sec:exp}

\subsection{Setup}
\label{sec:exp:set}

\paragraph{Hardware.} We implement our encoder and decoder units in SystemVerilog and use an EDA flow consisting of Synopsys and Cadence tools with the TSMC 7 nm FinFET technology to obtain latency and area.
We use Synopsys DesignWare library for a variety of RTL implementations such as the pipelined divider.
Power is estimated using Synopsys PrimeTimePX with fully annotated switching activity.

The DRAM energy is calculated using Micron's System Power
Calculators~\cite{microdrampower}, assuming a typical 8 Gb, 32-bit LPDDR4.
On average, the DRAM access energy per pixel is estimated to be 3,477 pJ/pixel, matching prior work~\cite{kodukula2021rhythmic, kodukula2021dynamic}.

\paragraph{Dataset and Software.} We evaluate our compression algorithm with 6 different VR scenes used in VR color perception studies~\cite{duinkharjav2022color}. In each scene, each frame is rendered with two sub-frames, one for each eye.
All the frames are dynamically rendered (on the GPU) at run time, i.e., the frames are neither loaded from the disk nor streamed over the network.
Following the common practice in color perception studies~\cite{cohen2020limits, duinkharjav2022color}, we keep pixels in the central 10\textdegree~FoV unchanged, and apply the compression algorithm only on the rest (peripheral) pixels.

As discussed in \Sect{sec:algo:algo}, our algorithm works in conjunction with existing BD compression.
In this paper, we assume a recent, state-of-the-art, BD algorithm described by Zhang et al.~\cite{zhang2019distilling}, from which we obtain the final compression rate.

\subsection{Human Subject Studies}
\label{sec:exp:human}

We also evaluate the perceptual quality of our compression algorithm on actual participants.
We recruit 11 participants (3 female; ages between 19 and 40).
None of the participants were aware of the research, the number of conditions, or the hypothesis before taking the experiments, which were approved by an Internal Review Board.

We face a dilemma in user study: the speed of the compression algorithm implemented as a GPU shader is too slow on today's mobile VR headsets (e.g., 2 FPS on Oculus Quest 2 as discussed in \Sect{sec:hw}) --- the motivation behind our architectural support, but this also means we can not use a mobile VR headset for user study.
Our approach is to run the user study on a tethered VR headset, HTC Vive Pro Eye, which is connected to a PC with a powerful Nvidia RTX A2000 GPU, which runs the compression algorithm at 90 FPS, sufficient for user study.

\begin{figure}[t]
\vspace{-15pt}
  \centering
    \subfloat[Original frame.]
    {
      \label{fig:2d}
      \includegraphics[width=.45\linewidth]{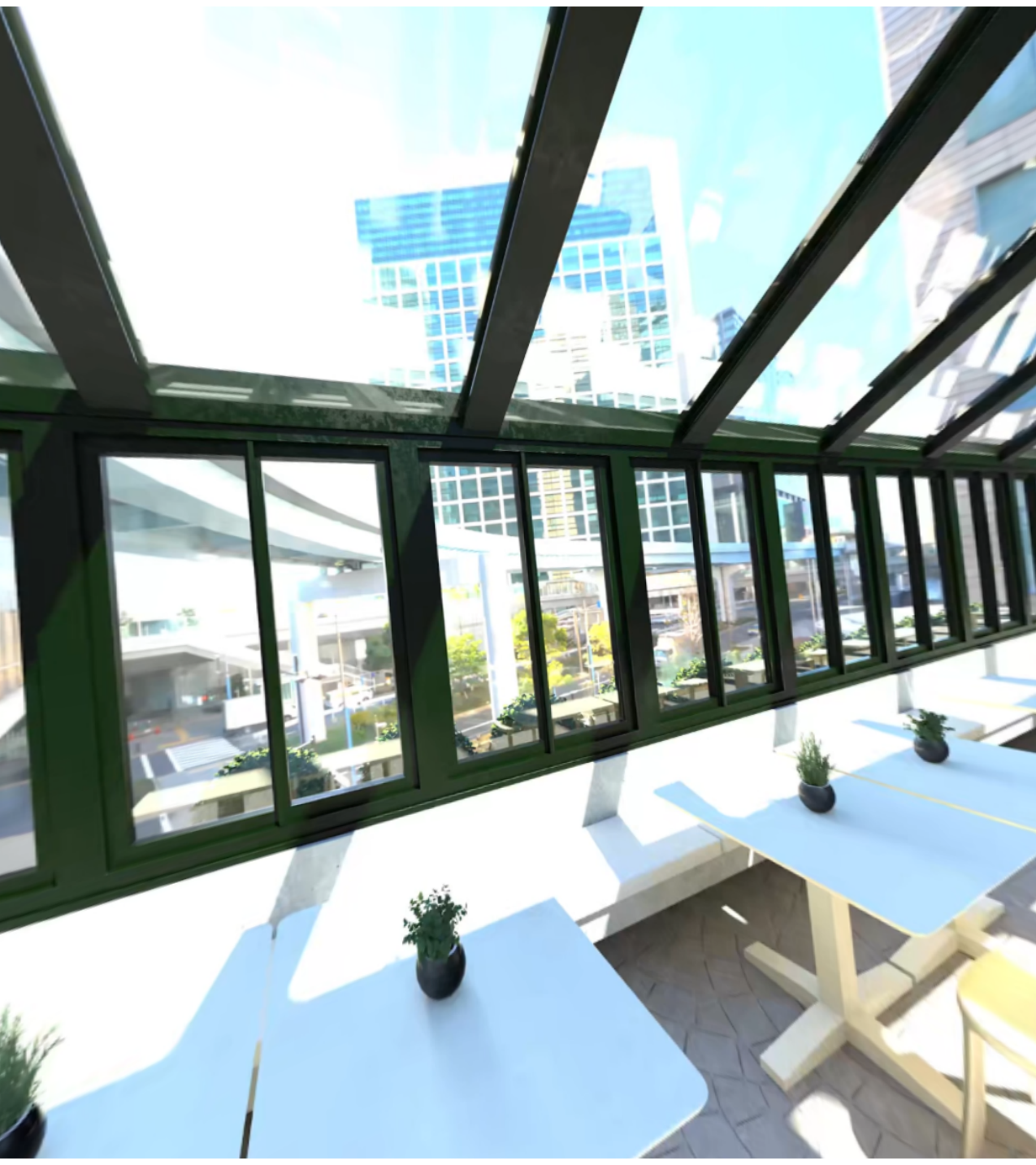}
    }
    \hfill
    \subfloat[Color-adjusted frame.]
    {
      \label{fig:2d-analog}
      \includegraphics[width=.45\linewidth]{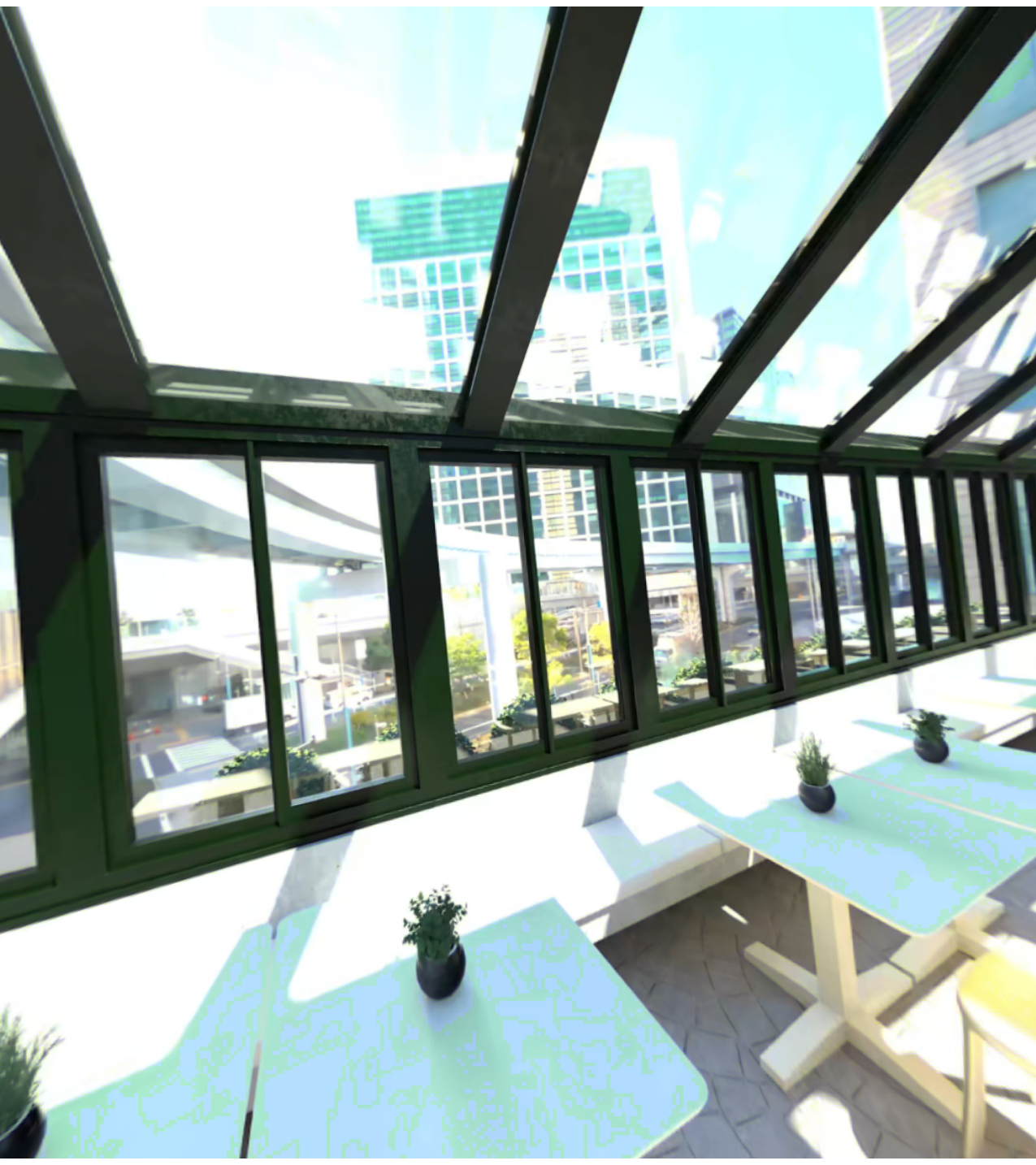}
    }
  \caption{A pair of images without (left) and with (right) our color adjustment. The two images when viewed on a conventional computer display are visibly different, because the entirety of the images will be in the viewer's foveal vision.}
\label{fig:example}
\end{figure}

Each participant was shown the six VR scenes (20 seconds each) used in a prior study~\cite{duinkharjav2022color} in random order.
To encourage and ensure that the participants actively and freely explored the scene, each participant was asked to perform a scene-specific task, such as counting the number of birds in the scene.
At the end of each video, we asked the participant whether they notice any visual artifacts.

In order for participants to isolate potential artifacts introduced by our compression from other irrelevant artifacts (e.g., low resolution, aliasing in rendering), at the beginning of each test we show the participant two images on a computer display, one with and the other without our perceptual compression; see examples in \Fig{fig:example}.
When participants viewed the images on the computer display, the entire frames were in their foveal vision so the color adjustment was clearly visible.
In this way, we make sure the artifacts reported by users resulted from compression.
This is a common practice in subjective color perception studies~\cite{duinkharjav2022color}.
The user study results should be seen as the \textit{lower bound} of the quality of compression algorithm, because the participants were aware of and thus better identification of the artifacts.

\subsection{Baselines}
\label{sec:exp:base}

We compare against four baselines:
\begin{itemize}
	\item \mode{NoCom}: no compression;
	\item \mode{BD}: existing BD compression based on Zhang et al.~\cite{zhang2019distilling};
	\item \mode{PNG}: lossless compression based on the popular Portable Network Graphics (PNG), which is unsuitable for real-time DRAM compression because of its high run-time overhead even with dedicated hardware acceleration~\cite{IPB-PNG-E, huang2008hardware}. For instance, the commercial IPB-PNG-E FPGA-based IP core compresses an $800 \times 600$ image only at a 20 FPS~\cite{IPB-PNG-E}.
	\item \mode{SCC}: an alternative strategy to exploit color discrimination based on the Set Cover formulation, which we describe next.
\end{itemize}

\mode{SCC} uses a look-up table to map each 24-bit sRGB color to a more compact encoding.
This can be formulated as a set cover problem~\cite{karp2010reducibility}: find the smallest subset of sRGB colors $\mathsf{C} \subset \mathsf{sRGB}$ whose discrimination ellipsoids union-ed together cover all the $2^{24}$ sRGB colors.
Each new color is then encoded with only $log_2\lceil |\mathsf{C}|\rceil$ bits, where $|\cdot|$ denotes the set cardinality.

The set cover problem is a classic NP-complete problem~\cite{karp2010reducibility}, where the optimal solution requires combinatorial search.
We use a common greedy heuristics~\cite{chvatal1979greedy} and construct the mapping tables.
The encoding table consumes 30 MB and the decoding table consumes 96 KB, too large for \mode{SCC} to be used for DRAM traffic compression in mobile SoCs.

%% file: sec_eval.tex
\section{Evaluation}
\label{sec:eval}

We first show that the area and power overhead of our compression scheme is negligible while ensuring real-time compression (\Sect{sec:eval:oh}).
We then present the benefits of our compression scheme in DRAM traffic reduction and power savings, and analyze the sources of the savings (\Sect{sec:eval:results}).
We then present our human subject studies, which show that our compression scheme introduces little visible artifacts (\Sect{sec:eval:hs}).
We present a sensitivity study of the key parameters in our compression scheme (\Sect{sec:eval:ss}). Finally, we discuss how we can accommodate a diverse range of users (\Sect{sec:eval:ds}).

\subsection{Area and Power Overhead}
\label{sec:eval:oh}

\paragraph{Performance.}
Our algorithm along with the hardware support achieves real-time compression.
The CAU operates with a cycle time of about 6~$ns$, which translates to a frequency of about 166.7~MHz.
The Adreno 650 GPU used in Oculus Quest 2 operates at a nominal frequency of 441 MHz, which means during each CAU cycle (at most) three pixels are generated by a shader core in the GPU.
Given that the Adreno 650 GPU has 512 shader cores, each CAU cycle $512 \times 3$ pixels (i.e., 96 tiles) are generated.
Therefore, we configure our CAU to have 96 PEs, which are able to process 96 tiles simultaneously, matching peak throughput of the GPU.

Thus, when compressing a 5408 $\times$ 2736 image (the highest rendering resolution on Oculus Quest 2), compression adds a delay of 173.4~$\mu s$, negligible in a rendering pipeline that operates at, say, 72 FPS with a frame time budget of 13.9~$ms$.


\paragraph{Area and Power.}
Our compression hardware extension introduces little area overhead, which consists of that of the Pending Buffers and the PEs.
Each PE has an area of 0.022 $mm^2$, resulting in a total PE size of 2.1 $mm^2$.
Each Pending Buffer holds data for two tiles (double buffering); the total buffer size is 36~KB, resulting in a total area of 0.03 $mm^2$.

The area overhead is negligible compared to the size of a typical mobile SoC. For instance, the Xavier SoC has an area of 350 $mm^2$ (12 nm)~\cite{xaviersoc}, Qualcomm Snapdragon 865 SoC has a die area of 83.54 $mm^2$ (7 nm)~\cite{snapdragon865}, and Apple A14 SoC has a die area of 88 $mm^2$ (5 nm)~\cite{applea14}.
The power consumption of each PE and its buffer is about 2.1 $\mu W$, resulting in a total CAU power consumption of about 201.6 $\mu W$, which we faithfully account for in the power saving analyses later.

\begin{figure}[t]
\vspace{2pt}
    \centering
    \includegraphics[width=\columnwidth]{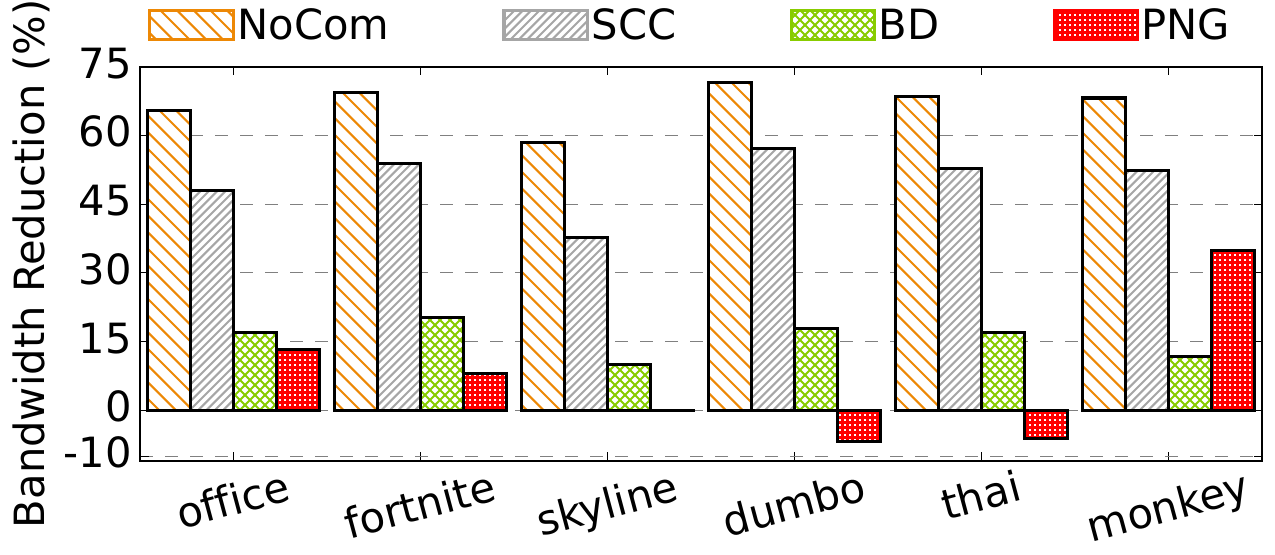}
    \caption{Bandwidth reduction over baselines.}
    \label{fig:comp}
\end{figure}

\begin{figure}[t]
    \centering
    \includegraphics[width=\columnwidth]{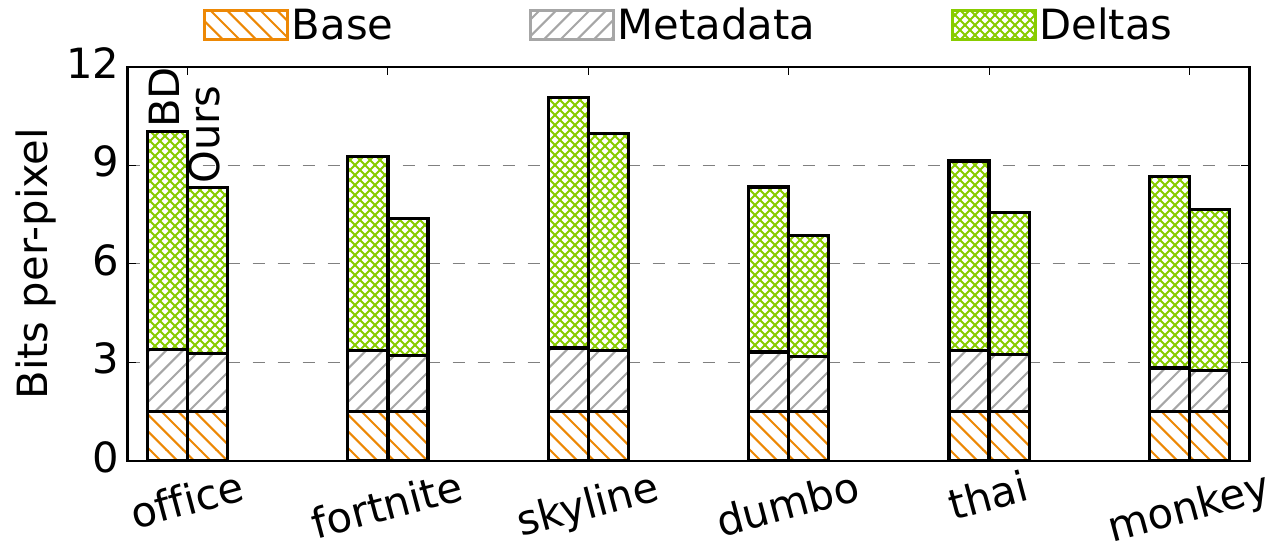}
    \caption{Distribution of bits per pixel across the three components: base, metadata, and $\Delta$. Left: BD; Right: our algorithm.}
    \label{fig:distribution}
\end{figure}

\subsection{Results}
\label{sec:eval:results}

\paragraph{Compression Rate.}
\Fig{fig:comp} shows the bandwidth reduction of our algorithm compared to the baselines.
Our algorithm achieves a compression rate of 66.9\%, 50.3\%, and 15.6\%  over \mode{NoCom}, \mode{SCC}, and \mode{BD}, respectively.
Unsurprisingly, the highest gains are against \mode{NoCom}, which is the original frames and uses 3 Bytes (24 bits) to store each pixel.

\mode{SCC} (\Sect{sec:exp:base}) is able to map all the $2^{24}$ (about 16.8 million) sRGB colors to a small subset of only 32,274 colors.
\mode{SCC} thus uses 15 bits to represent a color, reducing the storage cost compared to the original frames but is still much less efficient than \mode{BD}, which is the canonical Base+Delta approach to compression DRAM traffic in today's mobile SoCs.
Compared to \mode{BD}, we show 15.6\% (up to 20.4\%) higher compression rate, because of our ability to exploit human color discrimination to reduce the magnitudes of $\Delta$s.

We get the least improvement over \mode{PNG}.
In two scenes, \mode{PNG} actually has a higher compression rate.
This matches prior results on BD~\cite{zhang2019distilling} and is not surprising --- to get a high compression rate PNG is computationally intensive and is meant to be used offline; see discussion in \Sect{sec:exp:base}.

\paragraph{Understanding Results.}
Our improvement over BD comes from the fact that we require fewer bits to store the $\Delta$s.
\Fig{fig:distribution} shows the average number of bits per pixel required to store the base, metadata, and $\Delta$s in a tile.
We compare the statistics between BD (left bars) and our scheme (right bars).
It is clear that the space reduction comes from reducing the number of bits required to store the $\Delta$s.

\begin{figure}
    \centering
    \includegraphics[width=\columnwidth]{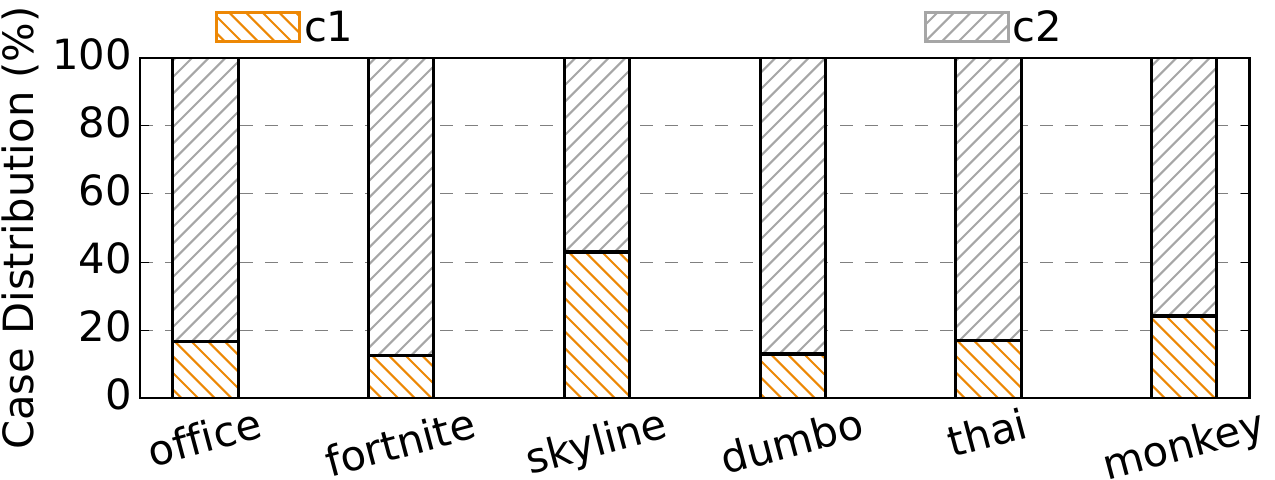}
    \caption{Distribution of the two cases \textsf{c1} and \textsf{c2}.}
    \label{fig:cases}
\end{figure}
To dissect how our scheme reduces the magnitude of $\Delta$s, \Fig{fig:cases} shows the distribution of tiles across the two cases in \Fig{fig:movc}: $\hhll > \llhh$ (\textsf{c1}) and $\hhll < \llhh$ (\textsf{c2}).
We observe that \textsf{c2} is the more common case:  78.92\% tiles result in this case.
In \textsf{c2}, there exists a common plane where all the color values can collapse to.
We can reduce the $\Delta$ to 0 in these tiles, essentially eliminating the need to store $\Delta$.

\begin{figure}[t]
    \centering
    \includegraphics[width=\linewidth]{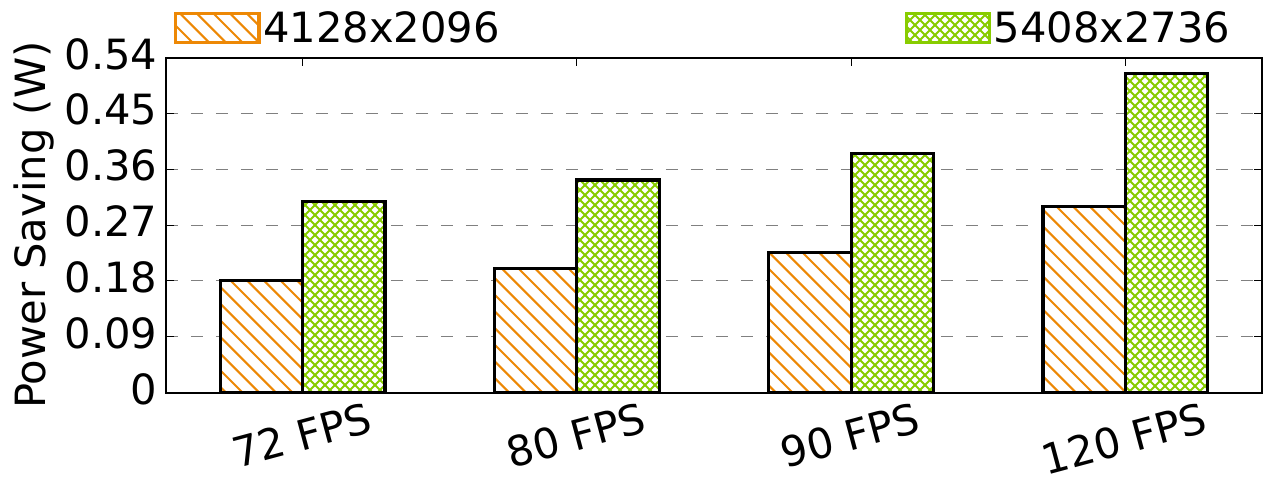}
    \caption{Power saving over \mode{BD} under the lowest and highest resolutions and four different frame rates on Oculus Quest 2.}
    \label{fig:power}
\end{figure}

\paragraph{Power Reduction.}
We evaluate the power reduction under different resolutions and frame rates available on Oculus Quest 2.
\Fig{fig:power} shows the power savings under each combination over \mode{BD}.
Across all configurations, we reduce the power consumption by 307.2 $mW$ on average.
The power saving is a combination of reducing the DRAM traffic and the power overhead of the CAU encoding (201.6 $\mu W$).

Even on the lowest resolution and frame rate combination on Oculus Quest 2, we reduce the power consumption by 180.3 $mW$, which translates to about 29.9\% of the total power measured (using Oculus' OVR Metrics Tool~\cite{ovrtool}) when rendering without compression.
Under the highest resolution and frame rate combination, the power saving increases to 514.2 $mW$.
As resolution and frame rate will likely increase in future VR devices, the power benefits of our compression scheme will only increase.

\subsection{User Studies and Analyses}
\label{sec:eval:hs}

\begin{figure}[t]
    \centering
    \includegraphics[width=\columnwidth]{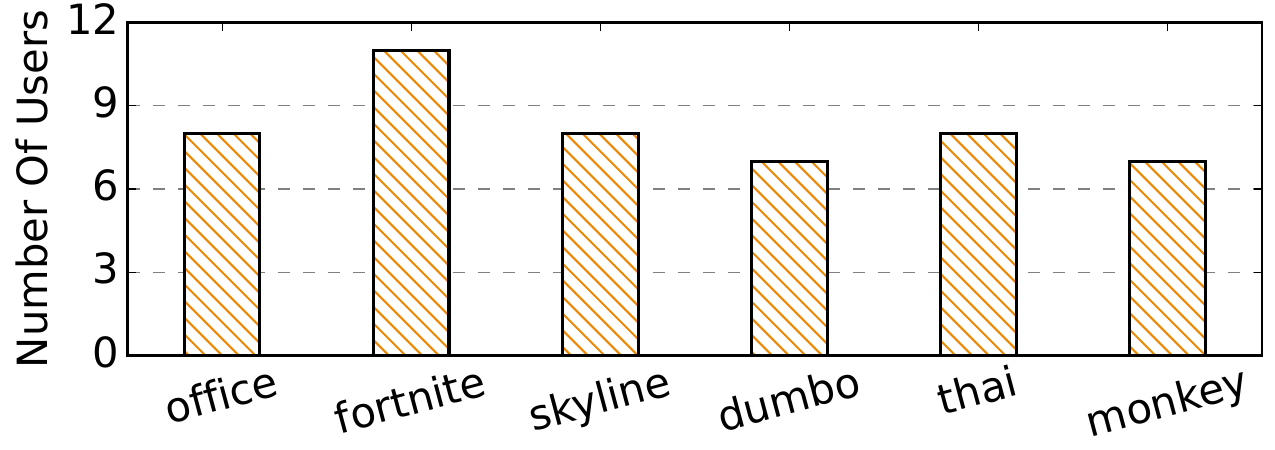}
    \caption{Number of participants (out of 11) who did \textit{not} notice any artifacts in each scene in our user study.}
    \label{fig:hs}
\end{figure}

\Fig{fig:hs} shows the number of participants who did \textit{not} notice any artifact in each scene.
On average, 2.8 participants (standard deviation 1.5) out of 11 total participants observe artifacts.
This percentage is on par with prior color perception studies~\cite{duinkharjav2022color, cohen2020limits}.
We further interviewed the participants and identified three reasons why certain participants notice artifacts, all of which were orthogonal to the fundamental idea of this paper and actually point to optimization opportunities in \textit{other} parts of the system, which, when exploited, can be readily integrated into our work.

One participant who noticed subtle artifacts in three out of the six scenes was a visual artist with ``color-sensitive eyes.''
Observer variation is a known phenomenon in vision science since the early days of color science research~\cite{xie2020observer, wright1929re, guild1931colorimetric}.
Given that color discrimination models in the literature all target the average case in the population, the results indicate that customizing/calibrating the model for individual users is likely a promising approach to reduce the artifact.

Another set of participants noticed artifacts only during rapid eye/head movement but not with a steady pose.
This is likely attributed to external factors such as rendering lag or slow gaze detection, which is independent of our algorithm.

Finally, we found that no participant noticed any artifact in the \textsf{fortnite} scene, which is a bright scene with a large amount of green.
Since our compression algorithm generally yields green-hue shifts (see examples in \Fig{fig:example}), artifacts are less noticeable in scenes that are green to begin with.
In contrast, \textsf{dumbo} and \textsf{monkey}, both dark scenes, have the most noticeable artifacts.
The results suggest, to the vision science community, the need for improving the color discrimination models in low-luminance conditions.

\paragraph{Objective Image Quality.}
To show that subjective experience, which is the focus of our work, is \textit{not} equivalent to objective quality, we evaluate the Peak-Signal-to-Noise-Ratio (PSNR), a common objective quality metric, of all the compressed images.
On average, the PSNR of the compressed videos is 46.0 dB (standard deviation 19.5); all but two scenes have a PSNR below 37.
A PSNR value in this range usually indicates noticeable visual artifacts~\cite{cao2015social}, which is confirmed by our participants when they view the compressed images on a conventional display.
This result accentuates the crux of our work:
use human color perception in VR to guide a numerically lossy scheme (hence low PSNR) for higher compression rate with little subjective quality degradation.



\subsection{Sensitivity Studies}
\label{sec:eval:ss}

\begin{figure}[t!]
\vspace{2pt}
    \centering
    \includegraphics[width=\columnwidth]{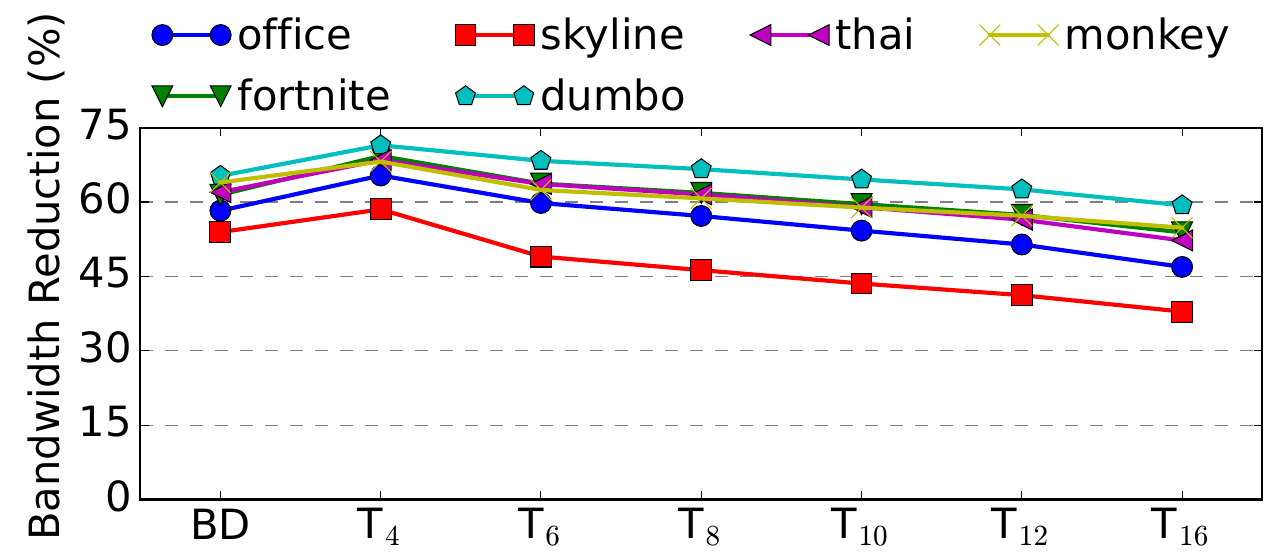}
    \caption{Bandwidth reduction over \mode{NoCom} under \mode{BD} and our scheme with different tile sizes denoted by $T_n$, where $n$ is the tile size.}
    \label{fig:tile}
\end{figure}

Our evaluation so far assumes a tile size of $4 \times 4$.
We also evaluate our compression algorithm across different tile sizes; the results are shown in \Fig{fig:tile} along with \mode{BD}.
We observe that the compression rate drops once the tile size increases beyond $4 \times 4$ and can be worse than \mode{BD} when the tile size is larger than $8 \times 8$.

The trend is the result of two opposing effects. On one hand, as we increase the tile size we can amortize the cost of storing the base pixels.
On the other hand, larger tiles also present less opportunity to bringing pixels together, because we have to accommodate the worst case/largest difference between two pixels in a tile (\Sect{sec:algo:ideas}).

\subsection{Discussions}
\label{sec:eval:ds}

To accommodate individual color perception in actual system deployments, one can perform a per-user color calibration procedure to build a per-user ellipsoid model. Such a procedure is laid out in prior work \mbox{\cite{duinkharjav2022color}}, and is readily doable. Such user-specific calibrations are routinely done when a user first uses an AR/VR product, e.g., adjusting the pair of displays to accommodate different inter-pupil distances among individuals. When a per-user ellipsoid model is available, our fundamental idea readily applies.

It is worth noting that we can, if need be, easily turn off our compression algorithm, which is intentionally designed as a plug-and-play stage between normal GPU rendering and existing BD compression (see \mbox{\Fig{fig:algo}}).
One scenario where one might want to turn off our compression is when a user has color vision deficiency (CVD). The color discrimination model that underlies our compression algorithm does not consider individuals with CVD. When such models for CVD become available, our fundamental idea readily applies.



%% file: sec_related.tex
\section{Related Work}
\label{sec:related}

\paragraph{Perception-Aware Rendering.}
A host of recent work has focused on leveraging human perception to optimize AR/VR systems. Weier et al. provide a relatively recent survey~\cite{weier2017perception}.
The most studied approach is foveated rendering, which reduces rendering resolution in the visual periphery~\cite{Patney:2016:TFR, Sun:2017:PGF, sun2020eccentricity, guenter2012foveated, walton2021beyond}.
Foveated rendering has been theoretically studied to reduce data transmission traffic in cloud rendering~\cite{Krajancich:2020:spatiotemp_model, kaplanyan2019deepfovea}, but the decoding (reconstruction) cost is prohibitively high (e.g., need a complicated DNN).
Our approach is orthogonal to foveated rendering in that we focus on adjusting colors rather than the spatial frequency, and works directly on top of the existing (BD-based) framebuffer compression framework without adding decoding cost.

\paragraph{Color Perception in Systems Optimizations.}
Color perception is most often leveraged to reduce display energy.
To our best knowledge, this is the first paper that leverages color perception to reduce data communication energy.

Dong et al. \cite{dong2009power}, Crayon~\cite{stanley2016crayon}, Dong and Zhong \cite{dong2011chameleon} all leverage the human color discrimination to reduce OLED power, which is known to strongly correlate with color.
Duinkharjav et al.~\cite{duinkharjav2022color} extend this approach to VR by quantifying the eccentricity dependent color discrimination.
Recent work by Dash and Hu \cite{dash2021much} builds an accurate color-vs-display power model.
None focused on reducing data traffic.
Shye et al.~\cite{shye2009into} and Yan et al.~\cite{yan2018exploring} leverage dark adaptation to reduce display power.
Dark adaptation will likely weaken the color discrimination even more, potentially further improving the compression rate --- an interesting future direction.

\paragraph{Data Traffic Optimizations in VR.}
Data traffic reduction in VR has mostly been studied in the context of client-cloud collaborative rendering, i.e., reducing wireless transmission traffic.
The pioneering Furion~\cite{lai2017furion} system and later developments and variants such as Coterie~\cite{meng2020coterie} and Q-VR~\cite{xie2021q} cleverly decide what to rendering locally vs. remotely.
For instance, one could offload background/far objects rendering to the server and render foreground/near object interactions locally.
EVR~\cite{leng2019energy, sun2020energy} predicts user FoV trajectory and pre-renders VR videos in the cloud.
Our proposal is orthogonal to the client-remote collaborative rendering, in that we focus on reducing DRAM traffic occurred within a local device.

Zhang et al.~\cite{zhang2019distilling} describe a BD design in encoding framebuffer traffic.
We directly compare against this approach and show up to 20\% bandwidth savings.
Zhang et al.~\cite{zhang2017race} propose a content cache that exploits value equality in video decoding, which does not apply to encoding where strict equality is rare.
Rhythmic Pixel Regions~\cite{kodukula2021rhythmic} drops pixel tiles to reduce DRAM traffic in a machine vision pipeline, whereas our focus is human visual perception in VR.

Any compression algorithm, ours included, exploits data similarities. Recent work leverages data similarities to speed-up rendering~\cite{zhao2020deja, zhao2021holoar, ying2022exploiting, wen2023post0} by eliding redundant computations (that compute same/similar data).
These methods, however, do not reduce data traffic, which we do.

\paragraph{General Memory Compression.}
Exploiting value similarities to compress data traffic is a long-standing technique in architecture~\cite{pekhimenko2012base, pekhimenko2013linearly}.
Recent work in approximate computing extends compression to tasks that can tolerate slight data infidelity such as image processing~\cite{miguel2015doppelganger, san2016bunker} and texture mapping in rendering~\cite{sutherland2015texture, xie2018perception}.
In comparison, this paper performs a principled ``approximate compression'' by 1) using a rigorous human perception model derived from psychophysical experiments and 2) formulating compression as a constraint optimization with an optimal solution (under necessary relaxations).
Finally, we specifically target VR and, thus, exploit the eccentricity dependency that is unconcerned with before.

%% file: sec_conc.tex
\section{Conclusion}
\label{sec:conc}

Aggressively lossy compression in the numerical domain can achieve significant data traffic reduction with little perceptual quality loss in VR.
The key is to leverage human color discrimination (in)ability to bring pixels more similar to each other.
The resulting images, thus, permit more aggressive compression over the classic Base+Delta scheme to reduce DRAM traffic in a mobile SoC.
We show that our compression algorithm has an analytical form, which, when accelerated by a dedicated hardware, can achieve real-time compression.
Future VR systems design must actively integrate human perception into the optimization loop.